\documentclass[10pt]{article}

\newcommand{\beq}{\begin{equation}}
\newcommand{\eeq}{\end{equation}}
\newcommand{\beqa}{\begin{eqnarray}}
\newcommand{\eeqa}{\end{eqnarray}}

\newcommand{\lmk}{\left(}
\newcommand{\rmk}{\right)}

\newcommand{\lnk}{\left\{}

\newcommand{\rhov}{\rho_{\rm v}}
\newcommand{\rhovz}{\rho_{{\rm v}0}}
\newcommand{\rhod}{\rho_{\rm d}}
\newcommand{\ns}{|n\rangle}
\newcommand{\thetas}{|\theta\rangle}

\newcommand{\cH}{{\cal H}}

\newcommand{\gtrsim}{~\mbox{\raisebox{-1.0ex}{$\stackrel{\textstyle >}
{\textstyle \sim}$ }}}
\newcommand{\lesssim}{~\mbox{\raisebox{-1.0ex}{$\stackrel{\textstyle <}
{\textstyle \sim}$ }}}
\begin{document}

\textwidth 11.9cm
\textheight 19cm
\voffset 0cm
\hoffset 0cm
%
%
\newcommand{\heading}[1]{\noindent{\large\bf{#1}}}
\renewenvironment{abstract}{\vskip 5mm
\noindent {\bf Abstract.\,}
  \renewcommand{\baselinestretch}{0.3}\small
}{}

\newcounter{adrc}\setcounter{adrc}{1}
\renewcommand{\author}[1]{\vskip 3mm \noindent{\rm #1}\vskip 3mm}
\newcommand{\address}[1]{\noindent$^{\theadrc}${\small\it #1}
\addtocounter{adrc}{1}\\}
%
%
%
\def\lsim{~\rlap{$&lt;$}{\lower 1.0ex\hbox{$\sim$}}}

\def\gsim{~\rlap{$&gt;$}{\lower 1.0ex\hbox{$\sim$}}}
%

%
\newcommand{\mn}{MNRAS\,\,}
\newcommand{\apj}{ApJ\,\,}
\newcommand{\aj}{AJ\,\,}
\newcommand{\aeta}{A\&amp;A\,\,}
\newcommand{\ass}{Astrophys. Space Sci.\,\,}
\newcommand{\nat}{Nature\,\,}
\newcommand{\et}{et al.\,\,}
\heading{%
%
Degenerate vacua as the origin of the dark energy
%
} 
\par\medskip\noindent
\author{%
Jun'ichi Yokoyama$^{1}$
}
\address{%
Department
of Earth and Space Science, Graduate School of Science,\\ Osaka
University, Toyonaka 560-0043, Japan
}

\begin{abstract}
I propose a new mechanism to account for the observed tiny but finite
dark energy in terms of a non-Abelian Higgs theory, which has
infinitely many perturbative vacua characterized by a winding number, 
in the framework of inflationary cosmology without introducing any 
tiny numbers.

\end{abstract}
\vskip 0.5cm
One of the greatest mysteries of present-day cosmology is the
origin of dark energy or vacuum-like component of cosmic energy
indicated by the analysis of type Ia supernovae
observations \cite{1}, whose magnitude is some 120 orders smaller
than its natural value,  the Planckian density \cite{4}.
I propose a novel 
mechanism to account for the observed tiny but finite
dark energy without introducing any small numbers
in the framework of inflationary cosmology \cite{5}
making use of a non-Abelian Higgs theory which is a part of
typical unified theories of elementary interactions.

Inflation   stretches  preexistent 
inhomogeneities
and yields an extremely large smooth domain in which the current Hubble
volume is contained.  It also provides a generation mechanism of 
density  fluctuations which originate in the quantum nature
of the inflation-driving scalar field \cite{6}.
In calculating  perturbations,
we consider the field to be in the vacuum state and calculate its
zero-point quantum fluctuations.  This treatment should be an
extremely good approximation, if not exact, as long as we focus on
perturbation within currently observable Hubble volume.  
This is because, even if the scalar field was in some excited
state initially, subsequent inflation would exponentially dilute the
preexistent quanta.  
Thus for all the practical purposes we can
regard that the quantum state of the scalar field is no different from
the vacuum state in this context.  That is to say, the vacuum state is
dynamically selected as a result of cosmological inflation.

Next let us consider a non-Abelian gauge theory with a gauge group $G$
which has infinitely many perturbative vacuum states classified by the 
winding number $n$.  Most gauge groups used in grand unified theories
such as $G=$SU($N$), SO($N$), and Sp($N$) with $N\geq 2$
possess this property.  For definiteness, however,
 we focus on the simplest theory with this property, $G=$SU(2), 
below.
In these theories perturbative vacuum 
states with different 
winding numbers cannot be transformed from each other by a continuous 
gauge transformation \cite{9} and there is an energy barrier between them.
Quantum transition, however, is possible  and 
described by an instanton solution which is an Euclidean solution
joining two adjacent perturbative vacua with a finite Euclidean
action $S_0=8\pi^2/g^2$. Here $g$ is the gauge coupling constant
 \cite{10}.  

As a result, the true vacuum state is expressed by a 
 superposition of perturbative vacua, $\ns$, as
\beq
  \thetas =\sum_{n=-\infty}^{\infty}e^{in\theta}\ns,
\eeq
where $\theta$ is a real parameter \cite{9,12}.
One can easily find that this state is a real
eigenstate of the Hamiltonian $\cH$ by 
calculating the amplitude 
$\left\langle {\theta '} \right|e^{-\cH T}\left| \theta  \right\rangle$
with the dilute instanton approximation \cite{12}, namely, adding
contributions of $m$ instantons and $\bar m$ anti-instantons keeping
in mind that each (anti-)instanton changes the winding number by $(-1)$1.
\beqa
\left\langle {\theta '} \right|e^{-\cH T}\left| \theta  \right\rangle
  &=&\sum\limits_{n,n'} {\sum\limits_{m,\bar m} {{1 \over {m! }}}{1
\over {\bar m! }}\left( {KVTe^{-S_0}} \right)^{m+\bar m}\delta _{m-\bar
m,n-n'}}e^{in\theta -in'\theta '}  \label{thetaex}  \\
  &=& \exp \left( {2KVTe^{-S_0}\cos \theta } \right)
\delta \left( {\theta -\theta '} \right). \nonumber 
\eeqa
Here $KVTe^{-S_0}$ represents  contribution of a single instanton or 
anti-instanton
where $K$ is a positive constant and
$VT$  represents spacetime volume.

The above equality (\ref{thetaex}) implies that each
$\theta$-vacuum $\thetas$ has a different energy density than the perturbative
vacuum $\ns$ by
$
\Delta\rho =-2Ke^{-S_0}\cos \theta . 
$
Apparently the $\theta=0$ vacuum has the lowest energy, 
but one cannot conclude that this is the only
vacuum state, because the other $\theta$-vacua are also stable against
gauge-invariant perturbations \cite{9,12}.  
In fact, the factor $K$  is  divergent due to the
contribution of arbitrary large instantons in pure gauge theory \cite{13}.  
In order to obtain a physical cutoff scale, let us introduce
an SU(2) doublet scalar field $\Phi$ with a potential 
$V[\Phi]=\lambda(|\Phi|^2-M^2/2)^2/2$ following 't Hooft \cite{14}.
For $M \neq 0$, instead of the instanton solution of pure gauge theory
we should use a constrained
instanton solution which also connects adjacent perturbative vacua with
a finite Euclidean action \cite{14}.  
Furthermore a gauge-invariant expression for the winding number can be
found in the presence of a Higgs field \cite{carena}.  
Then the prefactor is given
by \cite{16} $K\cong (8\pi/g^2)^4M^4/2$ and the expectation value of the
vacuum energy density in perturbative vacuum $\ns$ and that in
$\theta$-vacuum $\thetas$ are given by
\beqa
 \langle n|\cH\ns &=& \rhod + \rhov(\theta=0), \label{nenergy}\\
\rhov(\theta)\equiv \langle \theta|\cH\thetas &=& 
\rhod(1-\cos\theta)+\rhovz, 
\eeqa
respectively.  Here $\rhod\equiv (8\pi/g^2)^4M^4e^{-\frac{8\pi^2}{g^2}}$,
and $\rhov(\theta=0)\equiv\rhovz$ is the vacuum energy density in
the state $|\theta=0\rangle$.  Since different vacuum states have
different vacuum energy density, 
they behave differently in the presence of gravity.
Hence any good solution to the 
conventional cosmological constant problem,
namely why the vacuum energy density is vanishingly small compared with
the Planck scale, 
should specify at which vacuum state the cosmological constant vanishes.
Unfortunately, since there is no completely satisfactory solution to
this long-standing problem yet, let us simply 
normalize the vacuum energy density to vanish in the
CP-symmetric $\theta=0$ vacuum state and set $\rhovz=0$ for the moment just for
definiteness\footnote{ We  point out that the
wormhole mechanism \cite{17} is an example of proposed solutions to the
conventional cosmological constant problem which 
predicts the vacuum energy density vanishes at $\theta=0$ vacuum \cite{18}.}.  
As will be seen
below, however, this choice is not essential and  our mechanism
can work even if the vacuum energy density vanishes
at some other vacuum except at $|\theta=\pi\rangle$.

Now let us consider what quantum state is chosen by cosmological
inflation in this non-Abelian-Higgs system.  
Let us assume the energy scale of the Higgs field, $M$, is much larger
than the Hubble parameter during inflation, $H$, which is shown to be
the case later.
As in the case of singlet
scalar field, sufficiently long inflation would homogenize the Higgs 
field configuration over an exponentially large domain with $|\Phi|=M/\sqrt{2}$.
This in turn means that the state with vanishing winding number $|n=0\rangle$
is practically realized as long as we concentrate on the scales within the 
current Hubble volume, because this is the only state with homogeneous
scalar field configuration among many possible vacuum states, perturbative $\ns$
or real $\thetas$.  In terms of the energy eigenstates this state is
expanded as
\beq
 |n=0\rangle =\int_0^{2\pi}\frac{d\theta}{2\pi}\ \thetas , \label{zerovac}
\eeq 
which consists of superposition of all possible
real vacuum state $\thetas$ with equal weight.  

While we can
calculate the expectation value of vacuum energy density in this state
as (\ref{nenergy}), it is more appropriate to discuss the probability
distribution function (PDF) of vacuum energy because this state consists of
superposition of energy eigenstates with different energy density \cite{p}.
Using (\ref{zerovac}) we can easily calculate the PDF of vacuum energy
density, $P(\rhov)$, in this state as
\beqa
 P(\rhov)&=&\Bigl\langle \delta\Bigl( \rhov -\rhov(\theta)\Bigr)\Bigr\rangle_\theta
 =\int_0^{2\pi}\frac{d\theta}{2\pi}\delta\Bigl( \rhov
-\rhod(1-\cos\theta)\Bigr) \nonumber \\
&=&
\lnk\begin{array}{cl}
\frac{1}{\pi\lmk 2\rhov\rhod
-\rhov^2\rmk^{1/2}}& {\rm for}~0\leq \rhov \leq2\rhod ,
 \\
0 & {\rm otherwise.}  \\ 
\end{array}\right. \label{pdf}
\eeqa
It is sharply peaked at $\rhov =0$ corresponding to
$\theta=0$ and $\rhov=2\rhod$ corresponding to $\theta=\pi$ and the
probability to find $\rhov \approx 2\rhod$ is fairly large. 

Thus, if the perturbative vacuum $|n=0\rangle$ is stable in the
cosmological time scale, the PDF of the
vacuum energy density is still given by (\ref{pdf}) today 
in our universe which experienced inflation.
Hence it is by no means surprising that we
observe a finite dark energy around $2\rhod$ today.

We can easily match the predicted value, $\approx 2\rhod$, with the observed
one, $10^{-120}M_G^4$, by appropriately choosing $M$ and $g$, 
where $M_G$ is the reduced Planck scale.  We also demand 
that the tunneling rate from
$|n=0\rangle$, $\Gamma\approx Ke^{-\frac{16\pi^2}{g^2}}$, 
is small enough to guarantee that there is no
transition within the current Hubble volume, $H_0^{-3}$, in the cosmic age,
$\Gamma H^{-4}_0 \lesssim 1$.
Then we obtain
\beq
 {\pi}/{\alpha}+2\ln\alpha = 60\ln10
+2\ln\lmk{M}/{M_G}\rmk,~~~\mbox{and}~~~M \gtrsim  \alpha M_G,
\eeq
where $\alpha\equiv g^2/(4\pi)$ is the coupling strength at the energy
scale  $M/\sqrt{2}$.
If the above inequality  is marginally satisfied, we find
$\alpha=1/44.4$ and $M=5\times 10^{16}$GeV, which is much larger than 
the upper bound on both the amplitude of quantum fluctuation, $H/(2\pi)$,
during inflation \cite{gw} and the reheat temperature after 
inflation \cite{grav}.  
If, on the other hand, we
take $M=M_G$ so that the cutoff scale of instanton is identical to the
presumed field-theory cutoff, we find $\alpha=1/47.$

We note that it is not mandatory to assume that the vacuum energy
density vanishes at the state $|\theta=0\rangle$.  One may take
other vacuum state to normalize the origin of the vacuum energy density.
For example, if one assumes that $\rhov$ vanishes at $\ns$ or
$|\theta=\pi/2\rangle$, $P(\rhov)$
will be peaked at $\rhov=\pm\rhod$ instead, which we may interpret that
we live in a universe with $\rhov\approx\rhod$. 
Then the values of the model
parameters should only be slightly changed accordingly.

In conclusion, we have  reached a striking conclusion that 
the observed tiny dark energy can be realized by the
interplay of  high energy unified gauge theory and inflationary cosmology. 
Furthermore by virtue
of the exponential dependence of the cosmological constant on the action of
the instanton, we can reproduce its observed value without introducing any
tiny numbers.

%
\vfill
\end{document}